\documentclass[aps,pra,twocolumn, 10pt, showpacs,superscriptaddress,groupedaddress,floatfix]{revtex4-1} % for review and submission
\usepackage{xspace}
\usepackage[hidelinks]{hyperref}

\newcommand{\na}{\ensuremath{\mathbf\nabla}}

\newcommand{\ie}{\emph{i.e.} }
\newcommand{\eg}{\emph{e.g.}\@\xspace }

\usepackage{graphicx}
\usepackage{epstopdf}
\usepackage{setspace}
\usepackage{bbold}
\usepackage{comment}
\usepackage{color}
\usepackage{soul}
\usepackage[abs]{overpic}
\usepackage{amsmath}
\usepackage{lipsum}
\usepackage[abs]{overpic}
\usepackage{amsmath}
\usepackage{nicefrac}

\begin{document}

\author{Gadi Afek}
\affiliation{Wright Laboratory, Department of Physics, Yale University, New Haven, Connecticut 06520, USA}
\author{Fernando Monteiro}
\affiliation{Wright Laboratory, Department of Physics, Yale University, New Haven, Connecticut 06520, USA}
\author{Benjamin Siegel}
\affiliation{Wright Laboratory, Department of Physics, Yale University, New Haven, Connecticut 06520, USA}
\author{Jiaxiang Wang}
\affiliation{Wright Laboratory, Department of Physics, Yale University, New Haven, Connecticut 06520, USA}
\author{Sarah Dickson}
\affiliation{Wright Laboratory, Department of Physics, Yale University, New Haven, Connecticut 06520, USA}
\author{Juan Recoaro}
\affiliation{Wright Laboratory, Department of Physics, Yale University, New Haven, Connecticut 06520, USA}
\author{Molly Watts}
\affiliation{Wright Laboratory, Department of Physics, Yale University, New Haven, Connecticut 06520, USA}
\author{David C. Moore}
\affiliation{Wright Laboratory, Department of Physics, Yale University, New Haven, Connecticut 06520, USA}

\title[]{Control and measurement of electric dipole moments in levitated optomechanics}

\begin{abstract}
Levitated optomechanical systems are rapidly becoming leading tools for precision sensing, enabling a high level of control over the sensor's center of mass motion, rotation and electric charge state. Higher-order multipole moments in the charge distribution, however, remain a major source of backgrounds. By applying controlled precessive torques to the dipole moment of a levitated microsphere in vacuum, we demonstrate cancellation of dipole-induced backgrounds by 2 orders of magnitude. We measure the dipole moments of ng-mass spheres and determine their scaling with sphere size, finding that the dominant torques arise from induced dipole moments related to dielectric-loss properties of the SiO$_2$ spheres. Control of multipole moments in the charge distribution of levitated sensors is a key requirement to sufficiently reduce background sources in future applications.
\end{abstract}

\maketitle
%%%%%%%%%%%%%%%%%%%%%%%%%%%%%%%%%%%%%
%%%%%%%%%% INTRODUCTION %%%%%%%%%%%%%
%%%%%%%%%%%%%%%%%%%%%%%%%%%%%%%%%%%%%
\paragraph*{Introduction.}
Precision measurements utilizing the high sensitivity of optomechanical systems have become an important experimental tool over the past years. They have enabled groundbreaking tests of some of the most fundamental concepts in physics, such as the nature of gravity~\cite{LIGO2016,aggarwal2020challenges,Westphal2021,blakemore2021search} and electromagnetism~\cite{Parker2018,Afek2021}, and opened up new parameter space in the search for dark matter~\cite{Solaro2020,Counts2020,Monteiro2020DM}. Such systems are also at the forefront of exploring quantum mechanics at the macroscopic scale~\cite{Vinante2019,Delic2020,Tebbenjohanns2020}.

Key factors in the ability to achieve high force and acceleration sensitivities~\cite{Carney_2021,Moore_2021} and low temperatures~\cite{Delic2020,Tebbenjohanns2020,Monteiro2020uK} are the thermal and mechanical isolation and control over center of mass motion, rotational dynamics and electrical charge state of optically and magnetically trapped levitated objects in a high-vacuum environment. Typically, a feedback system is used to manipulate the mechanical degrees of freedom, whereas an electrical~\cite{Rider2019} or optical~\cite{Monteiro2018, Ahn2018GHz, Reimann2018GHz} control scheme is utilized to control the rotational degrees of freedom~\cite{Stickler_2021}. 

Controlling the electrical charge state of levitated objects is essential for many applications, since significant background forces can arise from coupling of the net charge of the sphere (or higher order multipole moments in its distribution) to stray electric fields. Previous work has demonstrated control over the net charge of levitated objects using a variety of techniques~\cite{Moore2014,Frimmer2017,Conangla:2018nnn,Monteiro2020uK,2020JPhD...53q5302B}. These techniques allow electrons to be selectively ejected either from the object itself, or from neighboring surfaces (to be eventually captured onto the object). By simultaneously measuring the motion of the object in an electric field, the net charge can be controlled with single electron precision.

Nullifying the net charge of the trapped object (\ie eliminating its electric monopole) reduces the most significant coupling to external electric fields. However, no technique demonstrated thus far has been capable of eliminating backgrounds related to higher multipole moments in the charge distribution. In the absence of net charge, the electric dipole moment typically provides the largest coupling to external electric fields and gradients, and is the leading contribution to the background for experiments such as the search for charge quantization and millicharged particles using levitated optomechanical sensors~\cite{Moore2014,Afek2021} and searches for new short-range interactions~\cite{Rider:2016_screened,blakemore2021search}. For ambitious future proposals employing levitated optomechanical sensors such as those aimed at detecting entanglement of two micron-sized masses using their mutual gravitational interaction~\cite{Marletto2017,Bose2017}, dipole-induced forces can be many orders-of-magnitude larger than the forces of interest. For example, for two levitated ng-mass spheres with a separation even as large as 100~$\mu$m~\cite{Marletto2017,Bose2017}, dipole-dipole interactions from a permanent dipole moment typical of existing measurements ($\approx 100~e\,\mu$m per sphere)~\cite{Rider2019,Rider:2016_screened} would induce a force 6 orders of magnitude larger than the desired gravitational interaction. Introduction of a conducting shield between the masses would be technically challenging and would not necessarily eliminate such forces due to the dipole-shield interaction~\cite{Garrett2020_casimir}. Thus, techniques to measure, control, and ultimately eliminate backgrounds related to multipole moments in the electric charge distribution of levitated objects are likely required if such applications are to be realized.

In this paper, we implement a method reminiscent of that used for many years in the fields of NMR~\cite{Hahn1950} and ultracold atoms~\cite{Andersen2003} to mitigate decoherence due to coupling to stray fields using controlled rotations of the spin-state. In this fully classical implementation, we apply electric fields and optical torques on a levitated sphere's dipole moment to induce controlled spatial rotations. We show that the sphere's center of mass response to an externally-applied electric field gradient flips phase as the sphere precesses around, reversing the effect of background forces arising from the field gradient. We further show that a sphere that is optically rotated at a high angular velocity will undergo oscillations in its angular acceleration as it decelerates and accelerates according to the direction of the precession. Additionally, measurements of the dipole moments for levitated SiO$_2$ spheres have thus far only been reported for $\lesssim 5~\mu$m-diameter spheres (with mass of $\sim100$~pg)~\cite{Rider2019}. Here we utilize our systems' unique capability to levitate larger, 1--10~ng-mass spheres and show that the permanent component of the dipole is roughly mass-independent whereas the induced term has a strong mass dependence, related to the dielectric-loss properties of the spheres. 

%%%%%%%%%%%%%%%%%%%%%%%%%%%%%%%%%%%%%%%%%%%%%%%%%%%%%%%%%%%%%%%%%%%%%%%%%%%%
%%%%%%%%%%%%%%%%%%%%%         Experimental     %%%%%%%%%%%%%%%%%%%%%%%%%%%%%
%%%%%%%%%%%%%%%%%%%%%%%%%%%%%%%%%%%%%%%%%%%%%%%%%%%%%%%%%%%%%%%%%%%%%%%%%%%%
\paragraph*{Setup.}

\begin{figure} %% Fig. 1, 
  \centering
    \begin{overpic}[width=\linewidth]{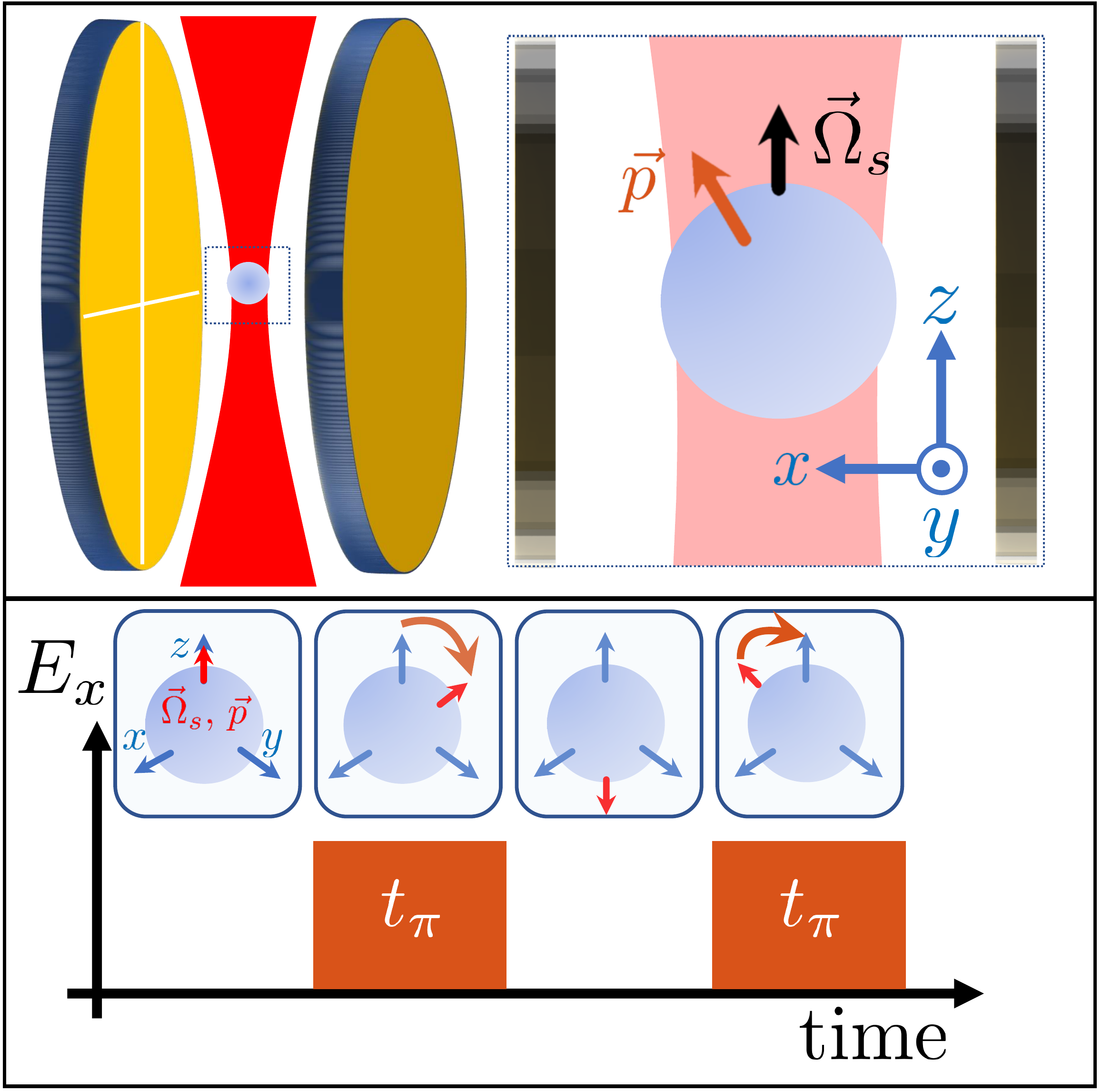}
    \put(2,115){\large \textbf{(a)}}
    \put(2,6){\large \textbf{(b)}}
    \end{overpic}
    \caption[]{Experimental setup and dipole control. (a) 10, 15 and 20~$\mu$m-diameter SiO$_2$ microspheres are optically levitated between a pair of parallel electrodes, one of which is segmented to enable application of arbitrary electric fields and gradients. The spheres can be spun optically, using a circularly polarized trapping beam, or electrically by coupling their electric dipole moment to a rotating electric field. (b) once the sphere reaches high angular frequencies a DC electric field can be applied to precess it, coupling to the remaining component of the dipole moment. One example is performing a sequence ``$\pi$-pulses", inverting the direction of the dipole vector and cancelling dipole-related backgrounds. The pulse length is given by $t_\pi = \pi/\Omega_p$ (Eq.~\ref{eq:precession})}
  \label{fig:fig1}
\end{figure}

In the experiment [Fig.~\ref{fig:fig1}~(a)], a vertically-oriented, 1064~nm laser beam is used to trap SiO$_2$ spheres of diameters $10, 15$ and 20~$\mu$m between a set of parallel electrodes 25.4~mm in diameter~\footnote{The spheres are grown chemically using the St$\ddot{\rm{o}}$ber process. See \url{https://www.microspheres-nanospheres.com/}}. The spheres are actively stabilized in high-vacuum and have typical trapping frequencies of $\sim100$~Hz~\cite{Monteiro2020uK}. One of the electrodes (left in Fig.~\ref{fig:fig1}) is divided into four separately-biased segments. The location of the sphere is calibrated by introducing onto it a net charge of $\sim100~e$ and recording its center-of-mass response in the $x$, $y$, and $z$ directions to a small oscillating electric field of $\sim1$~V/mm applied separately to each of the four electrodes. The ratio between these responses for a given axis is then compared to a COMSOL simulation of the trap geometry, and the position in space for which the correct response is obtained for all three axes is determined to be $(x,y,z) = (-0.1\pm0.1,0.38\pm0.02,-0.14\pm0.05)$~mm with respect to the center of the quadrant electrode (in $y$ and $z$) and half the distance between the opposite electrodes (in $x$). This position calibration allows calculation, for each electrode-segment $j$, of the voltage $V_j(t)$ required for the generation of a rotating electric field in the $x-y$ plane at the location of the sphere. The sphere can then be rotated either optically, via absorption and residual birefringence using a circularly-polarized trapping laser~\cite{Monteiro2018,Arita2013,Reimann2018GHz,Ahn2018GHz} or using the coupling of the rotating electric field to its dipole moment~\cite{Rider2019}. Rotations are measured using a polarization-sensitive detection scheme similar to the one described in~\cite{Monteiro2018}.

The force and torque on an electric dipole $\vec{p}$ in an electric field $\vec{E}$ are given, respectively, by $\vec{p}\cdot\na\vec{E}$ and $\vec{p}\times\vec{E}$. The total dipole moment $\vec{p} = \vec{p}_0 + \alpha\vec{E}$ can have a permanent component $\vec{p}_0$ and an induced component $\alpha\vec{E}$, brought about by the external electric field. Consider a spherical particle in vacuum, with radius $r$ and permittivity $\epsilon_s$, in an electric field $\vec{E} = \vec{E}_0 e^{i\omega t}$ oscillating at an angular frequency $\omega$. The polarizability can then be written as $\alpha=4\pi\epsilon_0 K r^3$, where $\epsilon_0$ is the vacuum permittivity. $K=(\epsilon_s^*-\epsilon_0)/(\epsilon_s^*+2\epsilon_0)$ is the Clausius-Mossotti factor for complex permitivitty $\epsilon_s^* = \epsilon_s + \sigma_s/(i\omega)$ and sphere conductivity $\sigma_s$~\cite{jones2005}. For a lossy dielectric (with finite, but non-zero, conductivity), $K$ is in general complex. The induced dipole moment can then lag after the field, generating induced torques that would vanish for either perfect conductors or ideal dielectrics. An aspherical particle will also have induced dipole moments that are not fully aligned with the applied field direction~\cite{jones2005}.

\paragraph*{Precessive control and background cancellation.}

Once the sphere is optically spun to a rotational speed $\Omega_s$ that is larger than other typical frequencies in the system, the components of the permanent dipole moment that are orthogonal to the spin axis are effectively averaged-out. A DC electric field $E$ in the $x$ direction precesses the sphere about the field direction at a frequency
\begin{equation}
    \Omega_p = \frac{pE}{I\Omega_s}.
    \label{eq:precession}
\end{equation}

Here $p=p_z$ is the component left after averaging out over the fast rotation. Cancellation of a background force arising from the coupling of the net dipole moment to external field gradients can be achieved by performing a measurement while the DC field is on and the sphere precesses continuously, or alternatively by applying a set of ``$\pi$-pulses" of duration $t_\pi = \pi/\Omega_p$ [Fig.~\ref{fig:fig1}~(b)] to invert the direction of the resultant force between measurements, as long as the timescale for the change of the background force is slow compared to the precession time. Dominant backgrounds in existing experiments are typically slow (\eg static stray fields). Attempting to mitigate those using static fields might result in elimination of a true signal. This is circumvented by controlling the alignment of the dipole itself, thus eliminating the need to know the details of the background field gradients. Furthermore, increasing the amplitude of the driving DC electric field will linearly increase the precession frequency and enable a faster pulse train. 

\begin{figure} %% Fig. 2, 
  \centering
    \begin{overpic}[width=\linewidth]{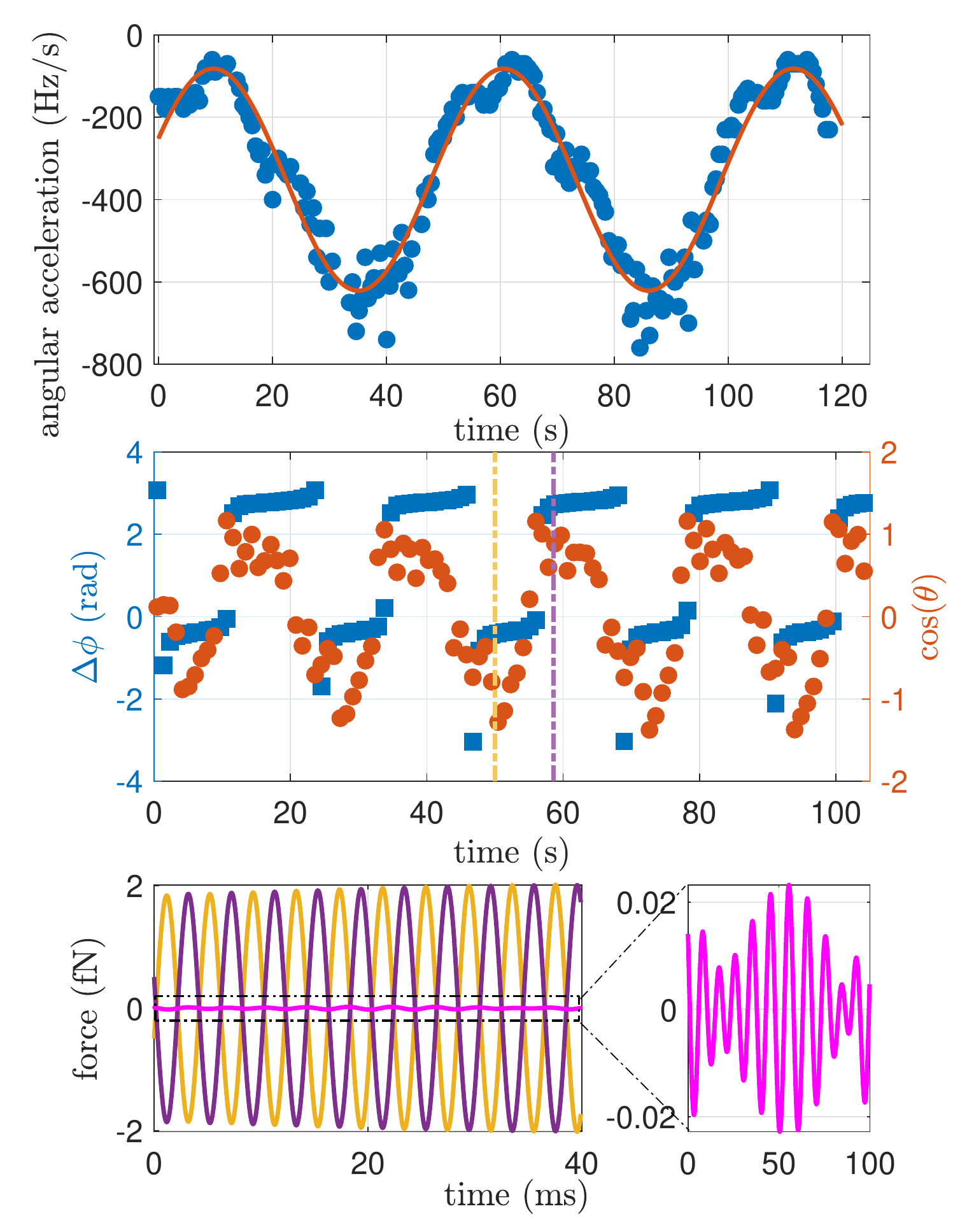}
    \put(42,230){\large \textbf{(a)}}
    \put(42,123){\large \textbf{(b)}}
    \put(42,30){\large \textbf{(c)}}
    \put(179,30){\large \textbf{(d)}}
    \end{overpic}
    \caption[]{Cancellation of backgrounds arising from dipole coupling to an externally applied electric field gradient. (a) Precession causes sinusoidal oscillations (blue circles, sine fit in solid red line) in the angular acceleration of the sphere, as the axis of $\Omega_s$ flips with respect to the constant optical torque. (b) The phase delay of the response of the sphere to a $2\pi\times 99$~Hz oscillating $\partial E_x/\partial z$ gradient (blue squares) exhibits a flip from in-phase to a $\pi$ phase lag, synchronous with the precession angle $\theta$ obtained from residual signal in the $y$ center-of-mass sensor (red circles). (c-d) The force acting on the sphere in the $x$ direction, sampled at two opposing points in the precession [purple (dark) and yellow (light), corresponding to the dash-dotted lines in (b)], bandpass-filtered to 10~Hz around the drive frequency. Over the 100~ms window, the sum of the two [magenta line surrounded by dash-dotted rectangle, zoomed-in in (d)] demonstrates cancellation by a factor of~$\sim 120$.}
  \label{fig:fig2}
\end{figure}
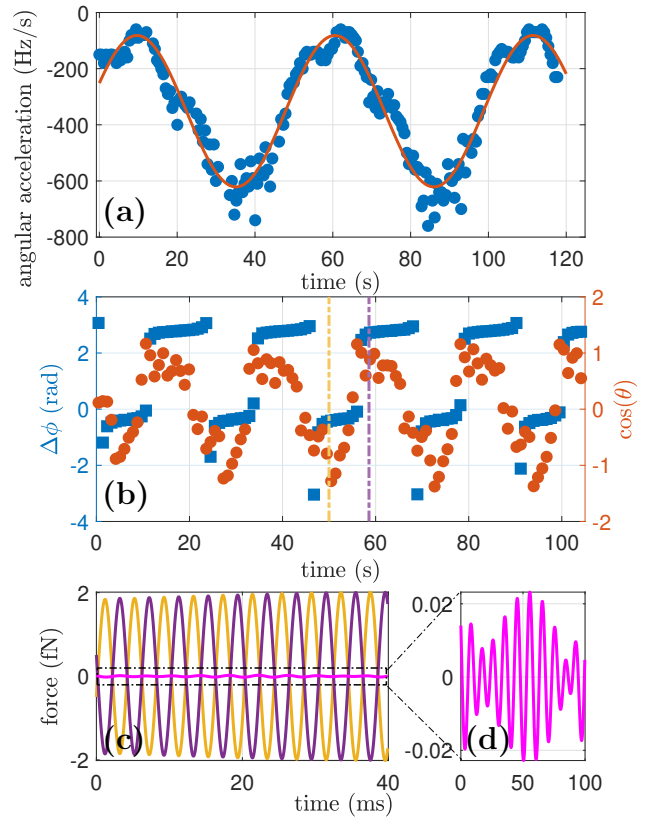

Fig.~\ref{fig:fig2} demonstrates this effect. A 15~$\mu$m-diameter sphere is pumped down to the $\sim 10^{-7}$~mbar base pressure of the system and optically spun up to $\Omega_s=2\pi\times1$~MHz and a 40~V/mm DC electric field is applied in the $x$ direction, generating a precession about the $x$ axis. Fig.~\ref{fig:fig2}~(a) shows this precession, measured via the change in the magnitude of the sphere's angular velocity over time. This data is obtained by tracking the peak of the rotational spectrum over $\sim1$~s increments and taking a numerical derivative. Since the optically-induced torque is always parallel to the $+z$ direction, manipulating the sign of $\Omega_s$ with respect to the $z$ axis results in the sphere's rotational velocity oscillating between slowing-down and speeding-up. An overall decelerating torque that is independent of the sphere orientation is also present, which may arise, \eg from drag from the background gas~\cite{Monteiro2018} or torques on the induced dipole from the applied electric field~\cite{jones2005}. Fitting these oscillations to a sine wave gives a precession frequency, according to Eq.~\ref{eq:precession}, of $2\pi\times(19.7\pm0.1)$~mHz.

The quadrant electrodes are then utilized, together with the simulated electric fields and gradients, to apply an AC electric field gradient at the location of the sphere, $\partial E_x/\partial z = 100$~V/mm$^2$. Constraints set by Maxwell's equations reduce the number of independent gradient components to five. Since the quadrant electrode provides only four controllable degrees of freedom, we optimize the drive to reduce the magnitude of the next largest parasitic gradient of $\partial E_y/\partial z\approx60$~V/mm$^2$ (generating a force in the $y$ direction which couples into the $x$ axis measurement with a negligible crosstalk of $<10\%$), and make all other independent components negligible. The sphere is electrically neutralized and hence residual forces due to the $<0.1$~V/mm electric fields (such as forces coupling to the dipole through geometrical gradients~\cite{Afek2021}) are negligible as well. The applied gradient couples to the $z$ component of the dipole to generate an oscillating force at the frequency of the drive, set to $\Omega_g = 2\pi\times 99$~Hz such that $\Omega_p\ll \Omega_g \ll \Omega_s$.  

The phase of the force acting on the sphere relative to the applied gradient shown in blue squares in Fig.~\ref{fig:fig2}~(b) as a function of time, as the sphere precesses under the influence of the DC field and in the presence of the oscillating gradient. The phase flips sharply between a value of $\sim0$ (in-phase with the drive) and $\sim\pi$ (out of phase with the drive). The deviation of the phase dynamics from a perfect step function can arise from other, sub-dominant precessive components in the motion of the sphere. The phase flip is synchronous with the angle $\theta$ between the angular momentum vector (which coincides with the effective dipole) and the $z$ axis. This angle is extracted from the center of mass sensors which are also sensitive to the double-frequency component of the angular motion due to the spheres' $\lesssim1\%$ inherent asphericity. This double-frequency sensitivity explains the factor of two difference in the measured precession frequency in (b) compared to (a). For two points in the process, corresponding to the sphere pointing at $\pm\hat{z}$ labelled by the yellow and purple dash-dotted lines in (b), we show in (c) and (d) a section of the $x$ force sensed by the sphere, bandpass-filtered to a 10~Hz band around the frequency of the drive to avoid harmonics and other noise lines. The magnitude of the force is consistent with a $\sim100~e\mu$m permanent dipole or a $\sim1\%$ asphericity-induced dipole in the applied field gradient. The sum of the forces measured over the chosen 100~ms integration window in these two points of the precessive motion is lower than the measured force by a factor of 120. This result demonstrates that, \eg a measurement protocol employing the $\pi$-pulse sequence described above would allow substantial mitigation of dipole-induced forces acting on the sphere by  averaging out asymmetries in the charge distribution of the sphere during the measurement time.

\paragraph*{Measurement of the dipole moment.}
Dipole moments are measured using three different techniques, previously demonstrated in~\cite{Rider2019} for smaller, $\sim100$~pg spheres. The first technique relies on measuring the harmonic librational motion about the axis of a rotating electric field of amplitude $E$. At the base pressure of our vacuum system, $\sim1\times10^{-7}$~mbar, phase lags resulting from damping by surrounding gas molecules are negligible, and the librational frequency $\Omega_L$ of a sphere with moment of inertia $I$ is given by
\begin{equation}
    \Omega_L = \sqrt{pE/I}.
    \label{eq:libration}
\end{equation}

The second technique employed to measure the dipole moment is via the precessive motion at a constant spin speed similar to that described in Eq.~\ref{eq:precession}, with $\Omega_s\gg\Omega_p,\Omega_L$. For precession about the rotating electric field, the expected precession frequency is $\Omega_p/2$, where the factor of 1/2 arises from the average torque over a full rotation of the microsphere~\cite{Rider2019}. Lastly, at high pressures of $\gtrsim 10^{-2}$~mbar, substantial phase lags between the dipole vector and the electric field can occur due to drag from the residual gas. When the phase lag becomes greater than $\pi/2$, the sphere will lose lock from the field and rapidly spin down. This occurs at a pressure $P_l$ and rotation frequency $\Omega_l$ related by
\begin{equation}
    \Omega_l = \frac{pE}{\beta\kappa_0P_l},
    \label{eq:lockloss}
\end{equation}
where $\kappa_0\approx 3.5\times10^{-23}$~m$^3$~s for a 15~$\mu$m diameter sphere (and scales with the fourth power of the diameter) is the proportionality constant relating the drag coefficient and the pressure~\cite{CAVALLERI20103365,PhysRevE.97.052112} and $\beta$ is a dimensionless quantity parameterizing the deviation of the actual drag from the theoretical value. Recent data from spheres with diameters $\lesssim 5~\mu$m have measured $\beta\approx1$, in agreement with its theoretical value~\cite{Rider2019}, whereas measurements of larger spheres, such as the ones used here, have obtained $\beta$ values of between 2 and 10, related to the surface quality of the spheres~\cite{Monteiro2018}.

\begin{figure} %% Fig. 3, 
  \centering
    \begin{overpic}[width=\linewidth]{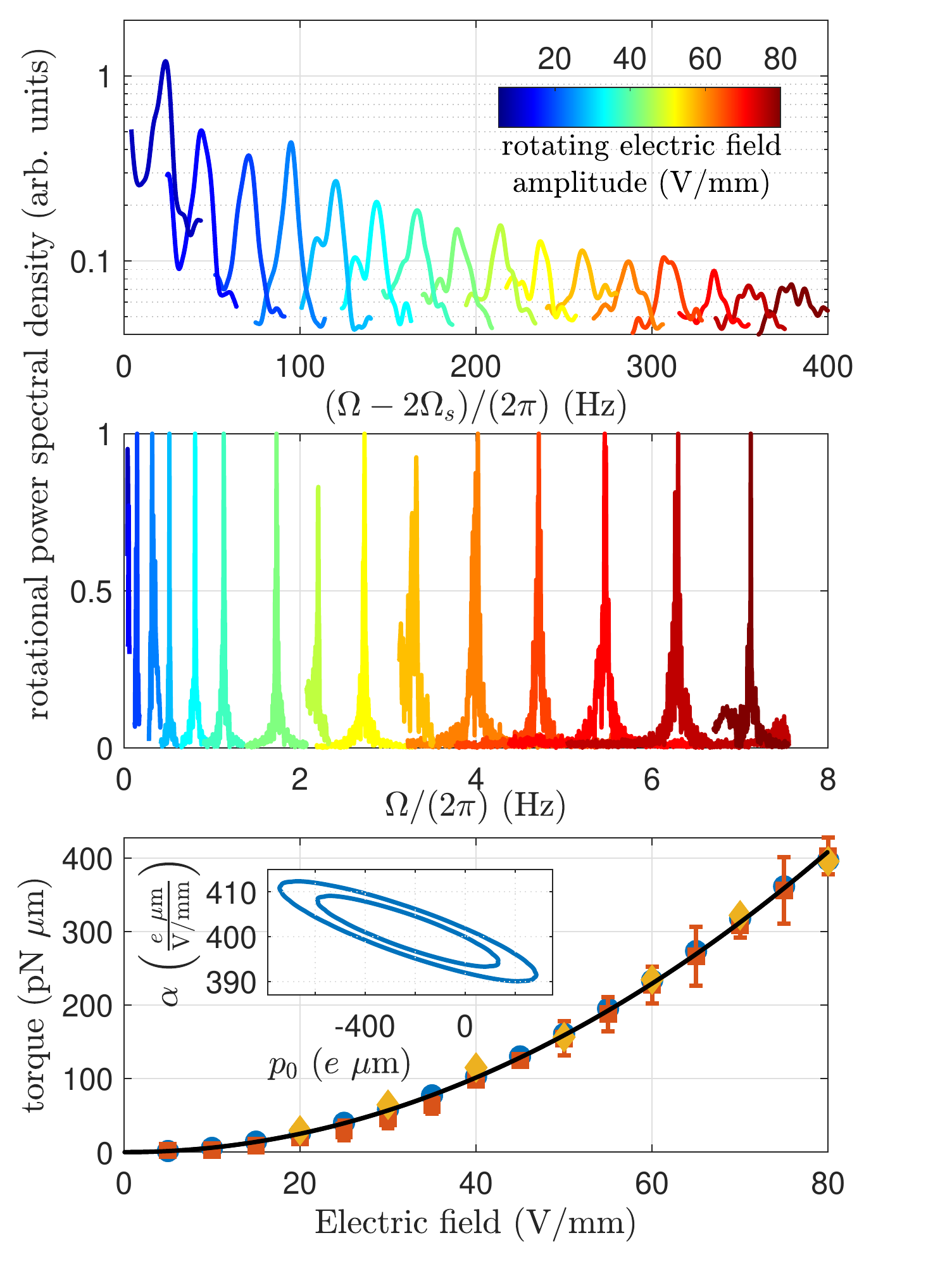}
    \put(35,250){\large \textbf{(a)}}
    \put(195,205){\large \textbf{(b)}}
    \put(195,38){\large \textbf{(c)}}
    \end{overpic}
    \caption[]{Measurement of the dipole moment using the precession and libration methods of Eq.~\ref{eq:precession} and \ref{eq:libration}. The sphere's rotation is measured in response to an electric field of varying amplitude (colorbar), rotating at a frequency $\Omega_s=2\pi\times10$~kHz. (a) Libration is observed as sidebands around the second harmonic $2\Omega_s$. (b) Precession is observed at lower frequencies. (c) Data from (a) and (b), rescaled according to Eq.~\ref{eq:libration}~(blue circles) and Eq.~\ref{eq:precession}~(red squares) respectively. Yellow diamonds represent high-pressure lock-loss measurements rescaled according to Eq.~\ref{eq:lockloss} with $\beta=6.97\pm0.07$. The inset shows the anti-correlated 1$\sigma$ and $2\sigma$ confidence interval contours.}
  \label{fig:fig3}
\end{figure}

Fig.~\ref{fig:fig3} presents a measurement of the electric dipole moment for a typical 15~$\mu$m sphere. For these measurements, the sphere is electrically spun at a constant $\Omega_s=2\pi\times 10$~kHz. The power spectral density of the signal recorded by the rotation sensor has two distinct sets of peaks. The first set, appearing predominantly as sidebands around the $2\Omega_s$ peak, corresponds to the librational motion of Eq.~\ref{eq:libration} [Fig.~\ref{fig:fig3}~(a)]~\cite{Rider2019}. The other set, at lower frequencies, corresponds to the precessive motion of Eq.~\ref{eq:precession} [Fig.~\ref{fig:fig3}~(b)]. Panel (c) shows the frequency values of the peaks from (a) in blue circles and (b) in red squares, rescaled to units of torque according to the appropriate equation. The overlap of the two data sets indicates the agreement between the methods. The yellow diamonds in (c) are the result of the lock-loss measurement at high pressure. Each data point is obtained, for a given pressure and rotating field amplitude, by scanning the rotation frequency of the field and monitoring the response of the sphere. The loss frequency is defined such that at $\Omega_s=\Omega_l$ the sphere spins rapidly down to zero. The results are rescaled using Eq.~\ref{eq:lockloss} and $\beta$ is allowed to float in the fit. The black line represents the best fit to the collapsed data obtained through a combined $\chi^2$ analysis of all three data sets after profiling over the value of $\beta$ as a nuisance parameter~\cite{ROLKE2005493}. The fit gives $\beta=6.97\pm0.07$ for this sphere and the resultant 1- and 2$\sigma$ confidence intervals on the fitted parameters $p_0$ and $\alpha$ are shown in the inset. The tilted confidence interval contour ellipse indicates that the parameters are anti-correlated. This is due to the fact that the total torque is the sum of the two, and the fit cannot completely differentiate between the permanent and induced components. Projecting the $1\sigma$ ellipse onto the respective axes gives the result $p_0 = (-200\pm360)~e\,\mu$m and $\alpha = 401\pm8~e\,\mu$m/(V/mm).

The primarily quadratic dependence of the torque on electric field indicates that the dominant dipole moment is induced by the electric field. As described above, such induced torques can arise for lossy dielectrics for which the imaginary part of the Clausius-Mossotti factor is significant. Asphericity is substantially smaller than would be required to explain the magnitude of the observed torques. The expected volume conductivity of pure SiO$_2$ is too small to explain the observed effect, although surface conductivity or lossy properties of the sol-gel SiO$_2$ spheres used here may account for the larger conductivity required to account for the observed torques.

\begin{figure} %% Fig. 4, 
  \centering
    \begin{overpic}[width=\linewidth]{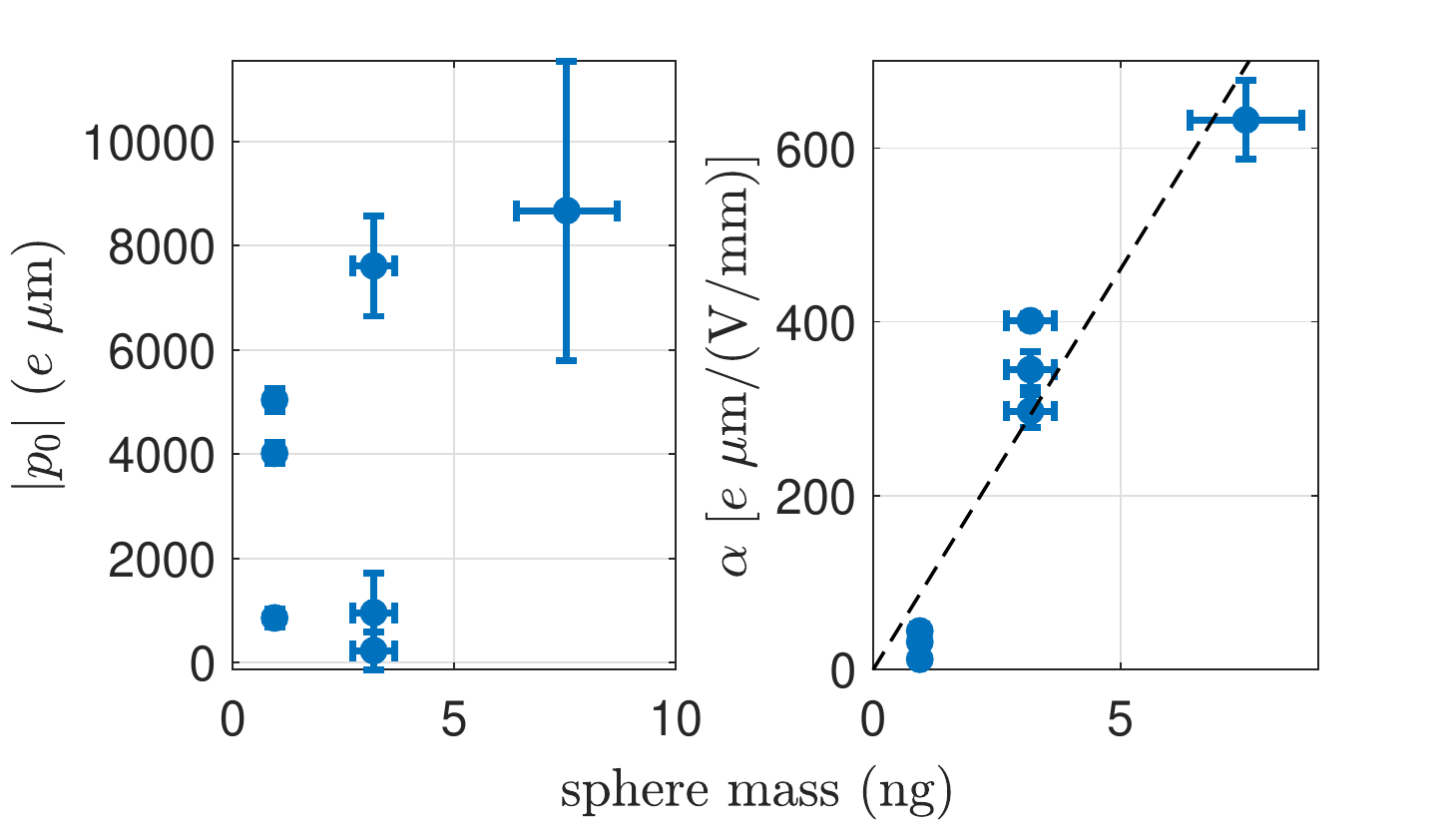}
    \put(95,35){\large \textbf{(a)}}
    \put(203,35){\large \textbf{(b)}}
    \end{overpic}
    \caption[]{Scaling of the dipole moment with sphere mass. 10, 15 and 20 $\mu$m diameter spheres are tested to compare the permanent (a) and induced (b) components of the total dipole. The dashed black line in (b) indicates the expected magnitude of $\alpha$ for a perfectly conducting sphere. }
  \label{fig:fig4}
\end{figure}

To study the dependence of the measured torques on sphere size, we apply the methods described above for spheres of several different diameters ($3\times10~\mu$m, $3\times15~\mu$m and $1\times20~\mu$m diameters). Fig.~\ref{fig:fig4} shows the permanent (a) and the induced (b) terms, extracted using a similar combined $\chi^2$ fit for libration and lock-loss datasets. Errors in $\vert p_0\vert$ and $\alpha$ come from the combined $\chi^2$ fit discussed in the text. Errors on mass result from a 10\% radius uncertainty. The permanent term, for which our measurement is less sensitive given the almost purely quadratic dependence of the total dipole on the electric field magnitude for the applied fields, does not exhibit significant mass-dependence. The induced term however scales with the sphere volume. For comparison, the dashed black line shows the value of $\alpha$ calculated for a perfectly conductive sphere of the same density as the examined spheres.

%%%%%%%%%%%%%%%%%%%%%%%%%%%%%%%%%%%%%%%%%%%%%%%%
%%%%%%%%%   Summary and outlook  %%%%%%
%%%%%%%%%%%%%%%%%%%%%%%%%%%%%%%%%%%%%%%%%%%%%%%%

\paragraph*{Summary and outlook.}
In conclusion, we have shown that controlled rotation and precession of an optically-trapped object can substantially assist in mitigating dipole-induced backgrounds. Combined with existing methods to control the net charge, the techniques presented here, classically analogous to the canonical spin-echo and dynamic decoupling methods used in the fields of NMR and ultracold atoms, can enable control over backgrounds related to dipole or higher order multipole moments in the charge distribution of a trapped object.

We have demonstrated that, in the objects studied here, significant torques arise from induced dipole moments which scale with the volume of the object and appear to arise from the lossy dielectric materials from which the SiO$_2$ spheres typically employed in levitated optomechanics are fabricated. Control of higher order multipole moments in the charge distribution of trapped particles may be required to reach sufficiently low environmental coupling in future applications of levitated optomechanical systems, such as searches for dark matter and new forces, hybrid and multi-particle systems, or attempts to witness gravitational entanglement between two levitated test masses.

\begin{acknowledgments}
The authors would like to thank Charles Blakemore and the Gratta group (Stanford) and Nir Davidson (Weizmann Institute) for discussions related to this work. This work is supported, in part, by ONR Grant N00014-18-1-2409, the Heising-Simons Foundation, and NSF Grant PHY-1653232.

\end{acknowledgments}

\bibliographystyle{apsrev4-1}
\bibliography{Dipoles}

\begin{thebibliography}{36}%
\makeatletter
\providecommand \@ifxundefined [1]{%
 \@ifx{#1\undefined}
}%
\providecommand \@ifnum [1]{%
 \ifnum #1\expandafter \@firstoftwo
 \else \expandafter \@secondoftwo
 \fi
}%
\providecommand \@ifx [1]{%
 \ifx #1\expandafter \@firstoftwo
 \else \expandafter \@secondoftwo
 \fi
}%
\providecommand \natexlab [1]{#1}%
\providecommand \enquote  [1]{``#1''}%
\providecommand \bibnamefont  [1]{#1}%
\providecommand \bibfnamefont [1]{#1}%
\providecommand \citenamefont [1]{#1}%
\providecommand \href@noop [0]{\@secondoftwo}%
\providecommand \href [0]{\begingroup \@sanitize@url \@href}%
\providecommand \@href[1]{\@@startlink{#1}\@@href}%
\providecommand \@@href[1]{\endgroup#1\@@endlink}%
\providecommand \@sanitize@url [0]{\catcode `\\12\catcode `\$12\catcode
  `\&12\catcode `\#12\catcode `\^12\catcode `\_12\catcode `\%12\relax}%
\providecommand \@@startlink[1]{}%
\providecommand \@@endlink[0]{}%
\providecommand \url  [0]{\begingroup\@sanitize@url \@url }%
\providecommand \@url [1]{\endgroup\@href {#1}{\urlprefix }}%
\providecommand \urlprefix  [0]{URL }%
\providecommand \Eprint [0]{\href }%
\providecommand \doibase [0]{http://dx.doi.org/}%
\providecommand \selectlanguage [0]{\@gobble}%
\providecommand \bibinfo  [0]{\@secondoftwo}%
\providecommand \bibfield  [0]{\@secondoftwo}%
\providecommand \translation [1]{[#1]}%
\providecommand \BibitemOpen [0]{}%
\providecommand \bibitemStop [0]{}%
\providecommand \bibitemNoStop [0]{.\EOS\space}%
\providecommand \EOS [0]{\spacefactor3000\relax}%
\providecommand \BibitemShut  [1]{\csname bibitem#1\endcsname}%
\let\auto@bib@innerbib\@empty
%</preamble>
\bibitem [{\citenamefont {Abbott}\ \emph {et~al.}(2016)\citenamefont {Abbott},
  \citenamefont {Abbott}, \citenamefont {Abbott}, \citenamefont {Abernathy},
  \citenamefont {Acernese}, \citenamefont {Ackley}, \citenamefont {Adams},
  \citenamefont {Adams}, \citenamefont {Addesso}, \citenamefont {Adhikari}
  \emph {et~al.}}]{LIGO2016}%
  \BibitemOpen
  \bibfield  {author} {\bibinfo {author} {\bibfnamefont {B.~P.}\ \bibnamefont
  {Abbott}}, \bibinfo {author} {\bibfnamefont {R.}~\bibnamefont {Abbott}},
  \bibinfo {author} {\bibfnamefont {T.}~\bibnamefont {Abbott}}, \bibinfo
  {author} {\bibfnamefont {M.}~\bibnamefont {Abernathy}}, \bibinfo {author}
  {\bibfnamefont {F.}~\bibnamefont {Acernese}}, \bibinfo {author}
  {\bibfnamefont {K.}~\bibnamefont {Ackley}}, \bibinfo {author} {\bibfnamefont
  {C.}~\bibnamefont {Adams}}, \bibinfo {author} {\bibfnamefont
  {T.}~\bibnamefont {Adams}}, \bibinfo {author} {\bibfnamefont
  {P.}~\bibnamefont {Addesso}}, \bibinfo {author} {\bibfnamefont
  {R.}~\bibnamefont {Adhikari}},  \emph {et~al.} (\bibinfo {collaboration}
  {LIGO Scientific Collaboration and Virgo Collaboration}),\ }\href {\doibase
  10.1103/PhysRevLett.116.061102} {\bibfield  {journal} {\bibinfo  {journal}
  {Physical review letters}\ }\textbf {\bibinfo {volume} {116}},\ \bibinfo
  {pages} {061102} (\bibinfo {year} {2016})}\BibitemShut {NoStop}%
\bibitem [{\citenamefont {Aggarwal}\ \emph {et~al.}(2020)\citenamefont
  {Aggarwal}, \citenamefont {Aguiar}, \citenamefont {Bauswein}, \citenamefont
  {Cella}, \citenamefont {Clesse}, \citenamefont {Cruise}, \citenamefont
  {Domcke}, \citenamefont {Figueroa}, \citenamefont {Geraci}, \citenamefont
  {Goryachev} \emph {et~al.}}]{aggarwal2020challenges}%
  \BibitemOpen
  \bibfield  {author} {\bibinfo {author} {\bibfnamefont {N.}~\bibnamefont
  {Aggarwal}}, \bibinfo {author} {\bibfnamefont {O.}~\bibnamefont {Aguiar}},
  \bibinfo {author} {\bibfnamefont {A.}~\bibnamefont {Bauswein}}, \bibinfo
  {author} {\bibfnamefont {G.}~\bibnamefont {Cella}}, \bibinfo {author}
  {\bibfnamefont {S.}~\bibnamefont {Clesse}}, \bibinfo {author} {\bibfnamefont
  {A.}~\bibnamefont {Cruise}}, \bibinfo {author} {\bibfnamefont
  {V.}~\bibnamefont {Domcke}}, \bibinfo {author} {\bibfnamefont
  {D.}~\bibnamefont {Figueroa}}, \bibinfo {author} {\bibfnamefont
  {A.}~\bibnamefont {Geraci}}, \bibinfo {author} {\bibfnamefont
  {M.}~\bibnamefont {Goryachev}},  \emph {et~al.},\ }\href@noop {} {\bibfield
  {journal} {\bibinfo  {journal} {arXiv preprint arXiv:2011.12414}\ } (\bibinfo
  {year} {2020})}\BibitemShut {NoStop}%
\bibitem [{\citenamefont {Westphal}\ \emph {et~al.}(2021)\citenamefont
  {Westphal}, \citenamefont {Hepach}, \citenamefont {Pfaff},\ and\
  \citenamefont {Aspelmeyer}}]{Westphal2021}%
  \BibitemOpen
  \bibfield  {author} {\bibinfo {author} {\bibfnamefont {T.}~\bibnamefont
  {Westphal}}, \bibinfo {author} {\bibfnamefont {H.}~\bibnamefont {Hepach}},
  \bibinfo {author} {\bibfnamefont {J.}~\bibnamefont {Pfaff}}, \ and\ \bibinfo
  {author} {\bibfnamefont {M.}~\bibnamefont {Aspelmeyer}},\ }\href {\doibase
  10.1038/s41586-021-03250-7} {\bibfield  {journal} {\bibinfo  {journal}
  {Nature}\ }\textbf {\bibinfo {volume} {591}},\ \bibinfo {pages} {225}
  (\bibinfo {year} {2021})}\BibitemShut {NoStop}%
\bibitem [{\citenamefont {Blakemore}\ \emph {et~al.}(2021)\citenamefont
  {Blakemore}, \citenamefont {Fieguth}, \citenamefont {Kawasaki}, \citenamefont
  {Priel}, \citenamefont {Martin}, \citenamefont {Rider}, \citenamefont
  {Wang},\ and\ \citenamefont {Gratta}}]{blakemore2021search}%
  \BibitemOpen
  \bibfield  {author} {\bibinfo {author} {\bibfnamefont {C.~P.}\ \bibnamefont
  {Blakemore}}, \bibinfo {author} {\bibfnamefont {A.}~\bibnamefont {Fieguth}},
  \bibinfo {author} {\bibfnamefont {A.}~\bibnamefont {Kawasaki}}, \bibinfo
  {author} {\bibfnamefont {N.}~\bibnamefont {Priel}}, \bibinfo {author}
  {\bibfnamefont {D.}~\bibnamefont {Martin}}, \bibinfo {author} {\bibfnamefont
  {A.~D.}\ \bibnamefont {Rider}}, \bibinfo {author} {\bibfnamefont
  {Q.}~\bibnamefont {Wang}}, \ and\ \bibinfo {author} {\bibfnamefont
  {G.}~\bibnamefont {Gratta}},\ }\href {https://arxiv.org/abs/2102.06848v1}
  {\bibfield  {journal} {\bibinfo  {journal} {arXiv}\ } (\bibinfo {year}
  {2021})},\ \Eprint {http://arxiv.org/abs/2102.06848} {arXiv:2102.06848
  [hep-ex]} \BibitemShut {NoStop}%
\bibitem [{\citenamefont {Parker}\ \emph {et~al.}(2018)\citenamefont {Parker},
  \citenamefont {Yu}, \citenamefont {Zhong}, \citenamefont {Estey},\ and\
  \citenamefont {M{\"u}ller}}]{Parker2018}%
  \BibitemOpen
  \bibfield  {author} {\bibinfo {author} {\bibfnamefont {R.~H.}\ \bibnamefont
  {Parker}}, \bibinfo {author} {\bibfnamefont {C.}~\bibnamefont {Yu}}, \bibinfo
  {author} {\bibfnamefont {W.}~\bibnamefont {Zhong}}, \bibinfo {author}
  {\bibfnamefont {B.}~\bibnamefont {Estey}}, \ and\ \bibinfo {author}
  {\bibfnamefont {H.}~\bibnamefont {M{\"u}ller}},\ }\href {\doibase
  10.1126/science.aap7706} {\bibfield  {journal} {\bibinfo  {journal}
  {Science}\ }\textbf {\bibinfo {volume} {360}},\ \bibinfo {pages} {191}
  (\bibinfo {year} {2018})}\BibitemShut {NoStop}%
\bibitem [{\citenamefont {Afek}\ \emph {et~al.}(2021)\citenamefont {Afek},
  \citenamefont {Monteiro}, \citenamefont {Wang}, \citenamefont {Siegel},
  \citenamefont {Ghosh},\ and\ \citenamefont {Moore}}]{Afek2021}%
  \BibitemOpen
  \bibfield  {author} {\bibinfo {author} {\bibfnamefont {G.}~\bibnamefont
  {Afek}}, \bibinfo {author} {\bibfnamefont {F.}~\bibnamefont {Monteiro}},
  \bibinfo {author} {\bibfnamefont {J.}~\bibnamefont {Wang}}, \bibinfo {author}
  {\bibfnamefont {B.}~\bibnamefont {Siegel}}, \bibinfo {author} {\bibfnamefont
  {S.}~\bibnamefont {Ghosh}}, \ and\ \bibinfo {author} {\bibfnamefont {D.~C.}\
  \bibnamefont {Moore}},\ }\href {\doibase 10.1103/PhysRevD.104.012004}
  {\bibfield  {journal} {\bibinfo  {journal} {Phys. Rev. D}\ }\textbf {\bibinfo
  {volume} {104}},\ \bibinfo {pages} {012004} (\bibinfo {year}
  {2021})}\BibitemShut {NoStop}%
\bibitem [{\citenamefont {Solaro}\ \emph {et~al.}(2020)\citenamefont {Solaro},
  \citenamefont {Meyer}, \citenamefont {Fisher}, \citenamefont {Berengut},
  \citenamefont {Fuchs},\ and\ \citenamefont {Drewsen}}]{Solaro2020}%
  \BibitemOpen
  \bibfield  {author} {\bibinfo {author} {\bibfnamefont {C.}~\bibnamefont
  {Solaro}}, \bibinfo {author} {\bibfnamefont {S.}~\bibnamefont {Meyer}},
  \bibinfo {author} {\bibfnamefont {K.}~\bibnamefont {Fisher}}, \bibinfo
  {author} {\bibfnamefont {J.~C.}\ \bibnamefont {Berengut}}, \bibinfo {author}
  {\bibfnamefont {E.}~\bibnamefont {Fuchs}}, \ and\ \bibinfo {author}
  {\bibfnamefont {M.}~\bibnamefont {Drewsen}},\ }\href {\doibase
  10.1103/PhysRevLett.125.123003} {\bibfield  {journal} {\bibinfo  {journal}
  {Phys. Rev. Lett.}\ }\textbf {\bibinfo {volume} {125}},\ \bibinfo {pages}
  {123003} (\bibinfo {year} {2020})}\BibitemShut {NoStop}%
\bibitem [{\citenamefont {Counts}\ \emph {et~al.}(2020)\citenamefont {Counts},
  \citenamefont {Hur}, \citenamefont {Aude~Craik}, \citenamefont {Jeon},
  \citenamefont {Leung}, \citenamefont {Berengut}, \citenamefont {Geddes},
  \citenamefont {Kawasaki}, \citenamefont {Jhe},\ and\ \citenamefont
  {Vuleti\ifmmode~\acute{c}\else \'{c}\fi{}}}]{Counts2020}%
  \BibitemOpen
  \bibfield  {author} {\bibinfo {author} {\bibfnamefont {I.}~\bibnamefont
  {Counts}}, \bibinfo {author} {\bibfnamefont {J.}~\bibnamefont {Hur}},
  \bibinfo {author} {\bibfnamefont {D.~P.~L.}\ \bibnamefont {Aude~Craik}},
  \bibinfo {author} {\bibfnamefont {H.}~\bibnamefont {Jeon}}, \bibinfo {author}
  {\bibfnamefont {C.}~\bibnamefont {Leung}}, \bibinfo {author} {\bibfnamefont
  {J.~C.}\ \bibnamefont {Berengut}}, \bibinfo {author} {\bibfnamefont
  {A.}~\bibnamefont {Geddes}}, \bibinfo {author} {\bibfnamefont
  {A.}~\bibnamefont {Kawasaki}}, \bibinfo {author} {\bibfnamefont
  {W.}~\bibnamefont {Jhe}}, \ and\ \bibinfo {author} {\bibfnamefont
  {V.}~\bibnamefont {Vuleti\ifmmode~\acute{c}\else \'{c}\fi{}}},\ }\href
  {\doibase 10.1103/PhysRevLett.125.123002} {\bibfield  {journal} {\bibinfo
  {journal} {Phys. Rev. Lett.}\ }\textbf {\bibinfo {volume} {125}},\ \bibinfo
  {pages} {123002} (\bibinfo {year} {2020})}\BibitemShut {NoStop}%
\bibitem [{\citenamefont {Monteiro}\ \emph
  {et~al.}(2020{\natexlab{a}})\citenamefont {Monteiro}, \citenamefont {Afek},
  \citenamefont {Carney}, \citenamefont {Krnjaic}, \citenamefont {Wang},\ and\
  \citenamefont {Moore}}]{Monteiro2020DM}%
  \BibitemOpen
  \bibfield  {author} {\bibinfo {author} {\bibfnamefont {F.}~\bibnamefont
  {Monteiro}}, \bibinfo {author} {\bibfnamefont {G.}~\bibnamefont {Afek}},
  \bibinfo {author} {\bibfnamefont {D.}~\bibnamefont {Carney}}, \bibinfo
  {author} {\bibfnamefont {G.}~\bibnamefont {Krnjaic}}, \bibinfo {author}
  {\bibfnamefont {J.}~\bibnamefont {Wang}}, \ and\ \bibinfo {author}
  {\bibfnamefont {D.~C.}\ \bibnamefont {Moore}},\ }\href {\doibase
  10.1103/PhysRevLett.125.181102} {\bibfield  {journal} {\bibinfo  {journal}
  {Phys. Rev. Lett.}\ }\textbf {\bibinfo {volume} {125}},\ \bibinfo {pages}
  {181102} (\bibinfo {year} {2020}{\natexlab{a}})}\BibitemShut {NoStop}%
\bibitem [{\citenamefont {Vinante}\ \emph {et~al.}(2019)\citenamefont
  {Vinante}, \citenamefont {Pontin}, \citenamefont {Rashid}, \citenamefont
  {Toro\ifmmode~\check{s}\else \v{s}\fi{}}, \citenamefont {Barker},\ and\
  \citenamefont {Ulbricht}}]{Vinante2019}%
  \BibitemOpen
  \bibfield  {author} {\bibinfo {author} {\bibfnamefont {A.}~\bibnamefont
  {Vinante}}, \bibinfo {author} {\bibfnamefont {A.}~\bibnamefont {Pontin}},
  \bibinfo {author} {\bibfnamefont {M.}~\bibnamefont {Rashid}}, \bibinfo
  {author} {\bibfnamefont {M.}~\bibnamefont {Toro\ifmmode~\check{s}\else
  \v{s}\fi{}}}, \bibinfo {author} {\bibfnamefont {P.~F.}\ \bibnamefont
  {Barker}}, \ and\ \bibinfo {author} {\bibfnamefont {H.}~\bibnamefont
  {Ulbricht}},\ }\href {\doibase 10.1103/PhysRevA.100.012119} {\bibfield
  {journal} {\bibinfo  {journal} {Phys. Rev. A}\ }\textbf {\bibinfo {volume}
  {100}},\ \bibinfo {pages} {012119} (\bibinfo {year} {2019})}\BibitemShut
  {NoStop}%
\bibitem [{\citenamefont {Deli{\'c}}\ \emph {et~al.}(2020)\citenamefont
  {Deli{\'c}}, \citenamefont {Reisenbauer}, \citenamefont {Dare}, \citenamefont
  {Grass}, \citenamefont {Vuleti{\'c}}, \citenamefont {Kiesel},\ and\
  \citenamefont {Aspelmeyer}}]{Delic2020}%
  \BibitemOpen
  \bibfield  {author} {\bibinfo {author} {\bibfnamefont {U.}~\bibnamefont
  {Deli{\'c}}}, \bibinfo {author} {\bibfnamefont {M.}~\bibnamefont
  {Reisenbauer}}, \bibinfo {author} {\bibfnamefont {K.}~\bibnamefont {Dare}},
  \bibinfo {author} {\bibfnamefont {D.}~\bibnamefont {Grass}}, \bibinfo
  {author} {\bibfnamefont {V.}~\bibnamefont {Vuleti{\'c}}}, \bibinfo {author}
  {\bibfnamefont {N.}~\bibnamefont {Kiesel}}, \ and\ \bibinfo {author}
  {\bibfnamefont {M.}~\bibnamefont {Aspelmeyer}},\ }\href {\doibase
  10.1126/science.aba3993} {\bibfield  {journal} {\bibinfo  {journal}
  {Science}\ }\textbf {\bibinfo {volume} {367}},\ \bibinfo {pages} {892}
  (\bibinfo {year} {2020})}\BibitemShut {NoStop}%
\bibitem [{\citenamefont {Tebbenjohanns}\ \emph {et~al.}(2020)\citenamefont
  {Tebbenjohanns}, \citenamefont {Frimmer}, \citenamefont {Jain}, \citenamefont
  {Windey},\ and\ \citenamefont {Novotny}}]{Tebbenjohanns2020}%
  \BibitemOpen
  \bibfield  {author} {\bibinfo {author} {\bibfnamefont {F.}~\bibnamefont
  {Tebbenjohanns}}, \bibinfo {author} {\bibfnamefont {M.}~\bibnamefont
  {Frimmer}}, \bibinfo {author} {\bibfnamefont {V.}~\bibnamefont {Jain}},
  \bibinfo {author} {\bibfnamefont {D.}~\bibnamefont {Windey}}, \ and\ \bibinfo
  {author} {\bibfnamefont {L.}~\bibnamefont {Novotny}},\ }\href {\doibase
  10.1103/PhysRevLett.124.013603} {\bibfield  {journal} {\bibinfo  {journal}
  {Phys. Rev. Lett.}\ }\textbf {\bibinfo {volume} {124}},\ \bibinfo {pages}
  {013603} (\bibinfo {year} {2020})}\BibitemShut {NoStop}%
\bibitem [{\citenamefont {Carney}\ \emph {et~al.}(2021)\citenamefont {Carney},
  \citenamefont {Krnjaic}, \citenamefont {Moore}, \citenamefont {Regal},
  \citenamefont {Afek}, \citenamefont {Bhave}, \citenamefont {Brubaker},
  \citenamefont {Corbitt}, \citenamefont {Cripe}, \citenamefont {Crisosto},
  \citenamefont {Geraci}, \citenamefont {Ghosh}, \citenamefont {Harris},
  \citenamefont {Hook}, \citenamefont {Kolb}, \citenamefont {Kunjummen},
  \citenamefont {Lang}, \citenamefont {Li}, \citenamefont {Lin}, \citenamefont
  {Liu}, \citenamefont {Lykken}, \citenamefont {Magrini}, \citenamefont
  {Manley}, \citenamefont {Matsumoto}, \citenamefont {Monte}, \citenamefont
  {Monteiro}, \citenamefont {Purdy}, \citenamefont {Riedel}, \citenamefont
  {Singh}, \citenamefont {Singh}, \citenamefont {Sinha}, \citenamefont
  {Taylor}, \citenamefont {Qin}, \citenamefont {Wilson},\ and\ \citenamefont
  {Zhao}}]{Carney_2021}%
  \BibitemOpen
  \bibfield  {author} {\bibinfo {author} {\bibfnamefont {D.}~\bibnamefont
  {Carney}}, \bibinfo {author} {\bibfnamefont {G.}~\bibnamefont {Krnjaic}},
  \bibinfo {author} {\bibfnamefont {D.~C.}\ \bibnamefont {Moore}}, \bibinfo
  {author} {\bibfnamefont {C.~A.}\ \bibnamefont {Regal}}, \bibinfo {author}
  {\bibfnamefont {G.}~\bibnamefont {Afek}}, \bibinfo {author} {\bibfnamefont
  {S.}~\bibnamefont {Bhave}}, \bibinfo {author} {\bibfnamefont
  {B.}~\bibnamefont {Brubaker}}, \bibinfo {author} {\bibfnamefont
  {T.}~\bibnamefont {Corbitt}}, \bibinfo {author} {\bibfnamefont
  {J.}~\bibnamefont {Cripe}}, \bibinfo {author} {\bibfnamefont
  {N.}~\bibnamefont {Crisosto}}, \bibinfo {author} {\bibfnamefont
  {A.}~\bibnamefont {Geraci}}, \bibinfo {author} {\bibfnamefont
  {S.}~\bibnamefont {Ghosh}}, \bibinfo {author} {\bibfnamefont {J.~G.~E.}\
  \bibnamefont {Harris}}, \bibinfo {author} {\bibfnamefont {A.}~\bibnamefont
  {Hook}}, \bibinfo {author} {\bibfnamefont {E.~W.}\ \bibnamefont {Kolb}},
  \bibinfo {author} {\bibfnamefont {J.}~\bibnamefont {Kunjummen}}, \bibinfo
  {author} {\bibfnamefont {R.~F.}\ \bibnamefont {Lang}}, \bibinfo {author}
  {\bibfnamefont {T.}~\bibnamefont {Li}}, \bibinfo {author} {\bibfnamefont
  {T.}~\bibnamefont {Lin}}, \bibinfo {author} {\bibfnamefont {Z.}~\bibnamefont
  {Liu}}, \bibinfo {author} {\bibfnamefont {J.}~\bibnamefont {Lykken}},
  \bibinfo {author} {\bibfnamefont {L.}~\bibnamefont {Magrini}}, \bibinfo
  {author} {\bibfnamefont {J.}~\bibnamefont {Manley}}, \bibinfo {author}
  {\bibfnamefont {N.}~\bibnamefont {Matsumoto}}, \bibinfo {author}
  {\bibfnamefont {A.}~\bibnamefont {Monte}}, \bibinfo {author} {\bibfnamefont
  {F.}~\bibnamefont {Monteiro}}, \bibinfo {author} {\bibfnamefont
  {T.}~\bibnamefont {Purdy}}, \bibinfo {author} {\bibfnamefont {C.~J.}\
  \bibnamefont {Riedel}}, \bibinfo {author} {\bibfnamefont {R.}~\bibnamefont
  {Singh}}, \bibinfo {author} {\bibfnamefont {S.}~\bibnamefont {Singh}},
  \bibinfo {author} {\bibfnamefont {K.}~\bibnamefont {Sinha}}, \bibinfo
  {author} {\bibfnamefont {J.~M.}\ \bibnamefont {Taylor}}, \bibinfo {author}
  {\bibfnamefont {J.}~\bibnamefont {Qin}}, \bibinfo {author} {\bibfnamefont
  {D.~J.}\ \bibnamefont {Wilson}}, \ and\ \bibinfo {author} {\bibfnamefont
  {Y.}~\bibnamefont {Zhao}},\ }\href {\doibase 10.1088/2058-9565/abcfcd}
  {\bibfield  {journal} {\bibinfo  {journal} {Quantum Science and Technology}\
  }\textbf {\bibinfo {volume} {6}},\ \bibinfo {pages} {024002} (\bibinfo {year}
  {2021})}\BibitemShut {NoStop}%
\bibitem [{\citenamefont {Moore}\ and\ \citenamefont
  {Geraci}(2021)}]{Moore_2021}%
  \BibitemOpen
  \bibfield  {author} {\bibinfo {author} {\bibfnamefont {D.~C.}\ \bibnamefont
  {Moore}}\ and\ \bibinfo {author} {\bibfnamefont {A.~A.}\ \bibnamefont
  {Geraci}},\ }\href {\doibase 10.1088/2058-9565/abcf8a} {\bibfield  {journal}
  {\bibinfo  {journal} {Quantum Science and Technology}\ }\textbf {\bibinfo
  {volume} {6}},\ \bibinfo {pages} {014008} (\bibinfo {year}
  {2021})}\BibitemShut {NoStop}%
\bibitem [{\citenamefont {Monteiro}\ \emph
  {et~al.}(2020{\natexlab{b}})\citenamefont {Monteiro}, \citenamefont {Li},
  \citenamefont {Afek}, \citenamefont {Li}, \citenamefont {Mossman},\ and\
  \citenamefont {Moore}}]{Monteiro2020uK}%
  \BibitemOpen
  \bibfield  {author} {\bibinfo {author} {\bibfnamefont {F.}~\bibnamefont
  {Monteiro}}, \bibinfo {author} {\bibfnamefont {W.}~\bibnamefont {Li}},
  \bibinfo {author} {\bibfnamefont {G.}~\bibnamefont {Afek}}, \bibinfo {author}
  {\bibfnamefont {C.-l.}\ \bibnamefont {Li}}, \bibinfo {author} {\bibfnamefont
  {M.}~\bibnamefont {Mossman}}, \ and\ \bibinfo {author} {\bibfnamefont
  {D.~C.}\ \bibnamefont {Moore}},\ }\href {\doibase
  10.1103/PhysRevA.101.053835} {\bibfield  {journal} {\bibinfo  {journal}
  {Phys. Rev. A}\ }\textbf {\bibinfo {volume} {101}},\ \bibinfo {pages}
  {053835} (\bibinfo {year} {2020}{\natexlab{b}})}\BibitemShut {NoStop}%
\bibitem [{\citenamefont {Rider}\ \emph {et~al.}(2019)\citenamefont {Rider},
  \citenamefont {Blakemore}, \citenamefont {Kawasaki}, \citenamefont {Priel},
  \citenamefont {Roy},\ and\ \citenamefont {Gratta}}]{Rider2019}%
  \BibitemOpen
  \bibfield  {author} {\bibinfo {author} {\bibfnamefont {A.~D.}\ \bibnamefont
  {Rider}}, \bibinfo {author} {\bibfnamefont {C.~P.}\ \bibnamefont
  {Blakemore}}, \bibinfo {author} {\bibfnamefont {A.}~\bibnamefont {Kawasaki}},
  \bibinfo {author} {\bibfnamefont {N.}~\bibnamefont {Priel}}, \bibinfo
  {author} {\bibfnamefont {S.}~\bibnamefont {Roy}}, \ and\ \bibinfo {author}
  {\bibfnamefont {G.}~\bibnamefont {Gratta}},\ }\href {\doibase
  10.1103/PhysRevA.99.041802} {\bibfield  {journal} {\bibinfo  {journal} {Phys.
  Rev. A}\ }\textbf {\bibinfo {volume} {99}},\ \bibinfo {pages} {041802}
  (\bibinfo {year} {2019})}\BibitemShut {NoStop}%
\bibitem [{\citenamefont {Monteiro}\ \emph {et~al.}(2018)\citenamefont
  {Monteiro}, \citenamefont {Ghosh}, \citenamefont {van Assendelft},\ and\
  \citenamefont {Moore}}]{Monteiro2018}%
  \BibitemOpen
  \bibfield  {author} {\bibinfo {author} {\bibfnamefont {F.}~\bibnamefont
  {Monteiro}}, \bibinfo {author} {\bibfnamefont {S.}~\bibnamefont {Ghosh}},
  \bibinfo {author} {\bibfnamefont {E.~C.}\ \bibnamefont {van Assendelft}}, \
  and\ \bibinfo {author} {\bibfnamefont {D.~C.}\ \bibnamefont {Moore}},\ }\href
  {\doibase 10.1103/PhysRevA.97.051802} {\bibfield  {journal} {\bibinfo
  {journal} {Phys. Rev. A}\ }\textbf {\bibinfo {volume} {97}},\ \bibinfo
  {pages} {051802} (\bibinfo {year} {2018})}\BibitemShut {NoStop}%
\bibitem [{\citenamefont {Ahn}\ \emph {et~al.}(2018)\citenamefont {Ahn},
  \citenamefont {Xu}, \citenamefont {Bang}, \citenamefont {Deng}, \citenamefont
  {Hoang}, \citenamefont {Han}, \citenamefont {Ma},\ and\ \citenamefont
  {Li}}]{Ahn2018GHz}%
  \BibitemOpen
  \bibfield  {author} {\bibinfo {author} {\bibfnamefont {J.}~\bibnamefont
  {Ahn}}, \bibinfo {author} {\bibfnamefont {Z.}~\bibnamefont {Xu}}, \bibinfo
  {author} {\bibfnamefont {J.}~\bibnamefont {Bang}}, \bibinfo {author}
  {\bibfnamefont {Y.-H.}\ \bibnamefont {Deng}}, \bibinfo {author}
  {\bibfnamefont {T.~M.}\ \bibnamefont {Hoang}}, \bibinfo {author}
  {\bibfnamefont {Q.}~\bibnamefont {Han}}, \bibinfo {author} {\bibfnamefont
  {R.-M.}\ \bibnamefont {Ma}}, \ and\ \bibinfo {author} {\bibfnamefont
  {T.}~\bibnamefont {Li}},\ }\href {\doibase 10.1103/PhysRevLett.121.033603}
  {\bibfield  {journal} {\bibinfo  {journal} {Phys. Rev. Lett.}\ }\textbf
  {\bibinfo {volume} {121}},\ \bibinfo {pages} {033603} (\bibinfo {year}
  {2018})}\BibitemShut {NoStop}%
\bibitem [{\citenamefont {Reimann}\ \emph {et~al.}(2018)\citenamefont
  {Reimann}, \citenamefont {Doderer}, \citenamefont {Hebestreit}, \citenamefont
  {Diehl}, \citenamefont {Frimmer}, \citenamefont {Windey}, \citenamefont
  {Tebbenjohanns},\ and\ \citenamefont {Novotny}}]{Reimann2018GHz}%
  \BibitemOpen
  \bibfield  {author} {\bibinfo {author} {\bibfnamefont {R.}~\bibnamefont
  {Reimann}}, \bibinfo {author} {\bibfnamefont {M.}~\bibnamefont {Doderer}},
  \bibinfo {author} {\bibfnamefont {E.}~\bibnamefont {Hebestreit}}, \bibinfo
  {author} {\bibfnamefont {R.}~\bibnamefont {Diehl}}, \bibinfo {author}
  {\bibfnamefont {M.}~\bibnamefont {Frimmer}}, \bibinfo {author} {\bibfnamefont
  {D.}~\bibnamefont {Windey}}, \bibinfo {author} {\bibfnamefont
  {F.}~\bibnamefont {Tebbenjohanns}}, \ and\ \bibinfo {author} {\bibfnamefont
  {L.}~\bibnamefont {Novotny}},\ }\href {\doibase
  10.1103/PhysRevLett.121.033602} {\bibfield  {journal} {\bibinfo  {journal}
  {Phys. Rev. Lett.}\ }\textbf {\bibinfo {volume} {121}},\ \bibinfo {pages}
  {033602} (\bibinfo {year} {2018})}\BibitemShut {NoStop}%
\bibitem [{\citenamefont {Stickler}\ \emph {et~al.}(2021)\citenamefont
  {Stickler}, \citenamefont {Hornberger},\ and\ \citenamefont
  {Kim}}]{Stickler_2021}%
  \BibitemOpen
  \bibfield  {author} {\bibinfo {author} {\bibfnamefont {B.~A.}\ \bibnamefont
  {Stickler}}, \bibinfo {author} {\bibfnamefont {K.}~\bibnamefont
  {Hornberger}}, \ and\ \bibinfo {author} {\bibfnamefont {M.~S.}\ \bibnamefont
  {Kim}},\ }\href {\doibase 10.1038/s42254-021-00335-0} {\bibfield  {journal}
  {\bibinfo  {journal} {Nature Reviews Physics}\ } (\bibinfo {year} {2021}),\
  10.1038/s42254-021-00335-0}\BibitemShut {NoStop}%
\bibitem [{\citenamefont {Moore}\ \emph {et~al.}(2014)\citenamefont {Moore},
  \citenamefont {Rider},\ and\ \citenamefont {Gratta}}]{Moore2014}%
  \BibitemOpen
  \bibfield  {author} {\bibinfo {author} {\bibfnamefont {D.~C.}\ \bibnamefont
  {Moore}}, \bibinfo {author} {\bibfnamefont {A.~D.}\ \bibnamefont {Rider}}, \
  and\ \bibinfo {author} {\bibfnamefont {G.}~\bibnamefont {Gratta}},\ }\href
  {\doibase 10.1103/PhysRevLett.113.251801} {\bibfield  {journal} {\bibinfo
  {journal} {Phys. Rev. Lett.}\ }\textbf {\bibinfo {volume} {113}},\ \bibinfo
  {pages} {251801} (\bibinfo {year} {2014})}\BibitemShut {NoStop}%
\bibitem [{\citenamefont {Frimmer}\ \emph {et~al.}(2017)\citenamefont
  {Frimmer}, \citenamefont {Luszcz}, \citenamefont {Ferreiro}, \citenamefont
  {Jain}, \citenamefont {Hebestreit},\ and\ \citenamefont
  {Novotny}}]{Frimmer2017}%
  \BibitemOpen
  \bibfield  {author} {\bibinfo {author} {\bibfnamefont {M.}~\bibnamefont
  {Frimmer}}, \bibinfo {author} {\bibfnamefont {K.}~\bibnamefont {Luszcz}},
  \bibinfo {author} {\bibfnamefont {S.}~\bibnamefont {Ferreiro}}, \bibinfo
  {author} {\bibfnamefont {V.}~\bibnamefont {Jain}}, \bibinfo {author}
  {\bibfnamefont {E.}~\bibnamefont {Hebestreit}}, \ and\ \bibinfo {author}
  {\bibfnamefont {L.}~\bibnamefont {Novotny}},\ }\href {\doibase
  10.1103/PhysRevA.95.061801} {\bibfield  {journal} {\bibinfo  {journal} {Phys.
  Rev. A}\ }\textbf {\bibinfo {volume} {95}},\ \bibinfo {pages} {061801}
  (\bibinfo {year} {2017})}\BibitemShut {NoStop}%
\bibitem [{\citenamefont {Conangla}\ \emph {et~al.}(2019)\citenamefont
  {Conangla}, \citenamefont {Ricci}, \citenamefont {Cuairan}, \citenamefont
  {Schell}, \citenamefont {Meyer},\ and\ \citenamefont
  {Quidant}}]{Conangla:2018nnn}%
  \BibitemOpen
  \bibfield  {author} {\bibinfo {author} {\bibfnamefont {G.~P.}\ \bibnamefont
  {Conangla}}, \bibinfo {author} {\bibfnamefont {F.}~\bibnamefont {Ricci}},
  \bibinfo {author} {\bibfnamefont {M.~T.}\ \bibnamefont {Cuairan}}, \bibinfo
  {author} {\bibfnamefont {A.~W.}\ \bibnamefont {Schell}}, \bibinfo {author}
  {\bibfnamefont {N.}~\bibnamefont {Meyer}}, \ and\ \bibinfo {author}
  {\bibfnamefont {R.}~\bibnamefont {Quidant}},\ }\href {\doibase
  10.1103/PhysRevLett.122.223602} {\bibfield  {journal} {\bibinfo  {journal}
  {Phys. Rev. Lett.}\ }\textbf {\bibinfo {volume} {122}},\ \bibinfo {pages}
  {223602} (\bibinfo {year} {2019})},\ \Eprint
  {http://arxiv.org/abs/1901.00923} {arXiv:1901.00923 [physics.ins-det]}
  \BibitemShut {NoStop}%
\bibitem [{\citenamefont {{Bullier}}\ \emph {et~al.}(2020)\citenamefont
  {{Bullier}}, \citenamefont {{Pontin}},\ and\ \citenamefont
  {{Barker}}}]{2020JPhD...53q5302B}%
  \BibitemOpen
  \bibfield  {author} {\bibinfo {author} {\bibfnamefont {N.~P.}\ \bibnamefont
  {{Bullier}}}, \bibinfo {author} {\bibfnamefont {A.}~\bibnamefont {{Pontin}}},
  \ and\ \bibinfo {author} {\bibfnamefont {P.~F.}\ \bibnamefont {{Barker}}},\
  }\href {\doibase 10.1088/1361-6463/ab71a7} {\bibfield  {journal} {\bibinfo
  {journal} {Journal of Physics D Applied Physics}\ }\textbf {\bibinfo {volume}
  {53}},\ \bibinfo {eid} {175302} (\bibinfo {year} {2020})},\ \Eprint
  {http://arxiv.org/abs/1906.09580} {arXiv:1906.09580 [physics.app-ph]}
  \BibitemShut {NoStop}%
\bibitem [{\citenamefont {Rider}\ \emph {et~al.}(2016)\citenamefont {Rider},
  \citenamefont {Moore}, \citenamefont {Blakemore}, \citenamefont {Louis},
  \citenamefont {Lu},\ and\ \citenamefont {Gratta}}]{Rider:2016_screened}%
  \BibitemOpen
  \bibfield  {author} {\bibinfo {author} {\bibfnamefont {A.~D.}\ \bibnamefont
  {Rider}}, \bibinfo {author} {\bibfnamefont {D.~C.}\ \bibnamefont {Moore}},
  \bibinfo {author} {\bibfnamefont {C.~P.}\ \bibnamefont {Blakemore}}, \bibinfo
  {author} {\bibfnamefont {M.}~\bibnamefont {Louis}}, \bibinfo {author}
  {\bibfnamefont {M.}~\bibnamefont {Lu}}, \ and\ \bibinfo {author}
  {\bibfnamefont {G.}~\bibnamefont {Gratta}},\ }\href {\doibase
  10.1103/PhysRevLett.117.101101} {\bibfield  {journal} {\bibinfo  {journal}
  {Phys. Rev. Lett.}\ }\textbf {\bibinfo {volume} {117}},\ \bibinfo {pages}
  {101101} (\bibinfo {year} {2016})},\ \Eprint
  {http://arxiv.org/abs/1604.04908} {arXiv:1604.04908 [hep-ex]} \BibitemShut
  {NoStop}%
\bibitem [{\citenamefont {Marletto}\ and\ \citenamefont
  {Vedral}(2017)}]{Marletto2017}%
  \BibitemOpen
  \bibfield  {author} {\bibinfo {author} {\bibfnamefont {C.}~\bibnamefont
  {Marletto}}\ and\ \bibinfo {author} {\bibfnamefont {V.}~\bibnamefont
  {Vedral}},\ }\href {\doibase 10.1103/PhysRevLett.119.240402} {\bibfield
  {journal} {\bibinfo  {journal} {Phys. Rev. Lett.}\ }\textbf {\bibinfo
  {volume} {119}},\ \bibinfo {pages} {240402} (\bibinfo {year}
  {2017})}\BibitemShut {NoStop}%
\bibitem [{\citenamefont {Bose}\ \emph {et~al.}(2017)\citenamefont {Bose},
  \citenamefont {Mazumdar}, \citenamefont {Morley}, \citenamefont {Ulbricht},
  \citenamefont {Toro\ifmmode~\check{s}\else \v{s}\fi{}}, \citenamefont
  {Paternostro}, \citenamefont {Geraci}, \citenamefont {Barker}, \citenamefont
  {Kim},\ and\ \citenamefont {Milburn}}]{Bose2017}%
  \BibitemOpen
  \bibfield  {author} {\bibinfo {author} {\bibfnamefont {S.}~\bibnamefont
  {Bose}}, \bibinfo {author} {\bibfnamefont {A.}~\bibnamefont {Mazumdar}},
  \bibinfo {author} {\bibfnamefont {G.~W.}\ \bibnamefont {Morley}}, \bibinfo
  {author} {\bibfnamefont {H.}~\bibnamefont {Ulbricht}}, \bibinfo {author}
  {\bibfnamefont {M.}~\bibnamefont {Toro\ifmmode~\check{s}\else \v{s}\fi{}}},
  \bibinfo {author} {\bibfnamefont {M.}~\bibnamefont {Paternostro}}, \bibinfo
  {author} {\bibfnamefont {A.~A.}\ \bibnamefont {Geraci}}, \bibinfo {author}
  {\bibfnamefont {P.~F.}\ \bibnamefont {Barker}}, \bibinfo {author}
  {\bibfnamefont {M.~S.}\ \bibnamefont {Kim}}, \ and\ \bibinfo {author}
  {\bibfnamefont {G.}~\bibnamefont {Milburn}},\ }\href {\doibase
  10.1103/PhysRevLett.119.240401} {\bibfield  {journal} {\bibinfo  {journal}
  {Phys. Rev. Lett.}\ }\textbf {\bibinfo {volume} {119}},\ \bibinfo {pages}
  {240401} (\bibinfo {year} {2017})}\BibitemShut {NoStop}%
\bibitem [{\citenamefont {Garrett}\ \emph {et~al.}(2020)\citenamefont
  {Garrett}, \citenamefont {Kim},\ and\ \citenamefont
  {Munday}}]{Garrett2020_casimir}%
  \BibitemOpen
  \bibfield  {author} {\bibinfo {author} {\bibfnamefont {J.~L.}\ \bibnamefont
  {Garrett}}, \bibinfo {author} {\bibfnamefont {J.}~\bibnamefont {Kim}}, \ and\
  \bibinfo {author} {\bibfnamefont {J.~N.}\ \bibnamefont {Munday}},\ }\href
  {\doibase 10.1103/PhysRevResearch.2.023355} {\bibfield  {journal} {\bibinfo
  {journal} {Phys. Rev. Research}\ }\textbf {\bibinfo {volume} {2}},\ \bibinfo
  {pages} {023355} (\bibinfo {year} {2020})}\BibitemShut {NoStop}%
\bibitem [{\citenamefont {Hahn}(1950)}]{Hahn1950}%
  \BibitemOpen
  \bibfield  {author} {\bibinfo {author} {\bibfnamefont {E.~L.}\ \bibnamefont
  {Hahn}},\ }\href {\doibase 10.1103/PhysRev.80.580} {\bibfield  {journal}
  {\bibinfo  {journal} {Phys. Rev.}\ }\textbf {\bibinfo {volume} {80}},\
  \bibinfo {pages} {580} (\bibinfo {year} {1950})}\BibitemShut {NoStop}%
\bibitem [{\citenamefont {Andersen}\ \emph {et~al.}(2003)\citenamefont
  {Andersen}, \citenamefont {Kaplan},\ and\ \citenamefont
  {Davidson}}]{Andersen2003}%
  \BibitemOpen
  \bibfield  {author} {\bibinfo {author} {\bibfnamefont {M.~F.}\ \bibnamefont
  {Andersen}}, \bibinfo {author} {\bibfnamefont {A.}~\bibnamefont {Kaplan}}, \
  and\ \bibinfo {author} {\bibfnamefont {N.}~\bibnamefont {Davidson}},\ }\href
  {\doibase 10.1103/PhysRevLett.90.023001} {\bibfield  {journal} {\bibinfo
  {journal} {Phys. Rev. Lett.}\ }\textbf {\bibinfo {volume} {90}},\ \bibinfo
  {pages} {023001} (\bibinfo {year} {2003})}\BibitemShut {NoStop}%
\bibitem [{Note1()}]{Note1}%
  \BibitemOpen
  \bibinfo {note} {The spheres are grown chemically using the St$\protect \ddot
  {\protect \rm {o}}$ber process. See \protect \url
  {https://www.microspheres-nanospheres.com/}}\BibitemShut {NoStop}%
\bibitem [{\citenamefont {{Arita}}\ \emph {et~al.}(2013)\citenamefont
  {{Arita}}, \citenamefont {{Mazilu}},\ and\ \citenamefont
  {{Dholakia}}}]{Arita2013}%
  \BibitemOpen
  \bibfield  {author} {\bibinfo {author} {\bibfnamefont {Y.}~\bibnamefont
  {{Arita}}}, \bibinfo {author} {\bibfnamefont {M.}~\bibnamefont {{Mazilu}}}, \
  and\ \bibinfo {author} {\bibfnamefont {K.}~\bibnamefont {{Dholakia}}},\
  }\href {\doibase 10.1038/ncomms3374} {\bibfield  {journal} {\bibinfo
  {journal} {Nature Communications}\ }\textbf {\bibinfo {volume} {4}},\
  \bibinfo {eid} {2374} (\bibinfo {year} {2013})}\BibitemShut {NoStop}%
\bibitem [{\citenamefont {Jones}\ and\ \citenamefont
  {Jones}(2005)}]{jones2005}%
  \BibitemOpen
  \bibfield  {author} {\bibinfo {author} {\bibfnamefont {T.~B.}\ \bibnamefont
  {Jones}}\ and\ \bibinfo {author} {\bibfnamefont {T.~B.}\ \bibnamefont
  {Jones}},\ }\href@noop {} {\emph {\bibinfo {title} {Electromechanics of
  particles}}}\ (\bibinfo  {publisher} {Cambridge university press},\ \bibinfo
  {year} {2005})\BibitemShut {NoStop}%
\bibitem [{\citenamefont {Cavalleri}\ \emph {et~al.}(2010)\citenamefont
  {Cavalleri}, \citenamefont {Ciani}, \citenamefont {Dolesi}, \citenamefont
  {Hueller}, \citenamefont {Nicolodi}, \citenamefont {Tombolato}, \citenamefont
  {Vitale}, \citenamefont {Wass},\ and\ \citenamefont
  {Weber}}]{CAVALLERI20103365}%
  \BibitemOpen
  \bibfield  {author} {\bibinfo {author} {\bibfnamefont {A.}~\bibnamefont
  {Cavalleri}}, \bibinfo {author} {\bibfnamefont {G.}~\bibnamefont {Ciani}},
  \bibinfo {author} {\bibfnamefont {R.}~\bibnamefont {Dolesi}}, \bibinfo
  {author} {\bibfnamefont {M.}~\bibnamefont {Hueller}}, \bibinfo {author}
  {\bibfnamefont {D.}~\bibnamefont {Nicolodi}}, \bibinfo {author}
  {\bibfnamefont {D.}~\bibnamefont {Tombolato}}, \bibinfo {author}
  {\bibfnamefont {S.}~\bibnamefont {Vitale}}, \bibinfo {author} {\bibfnamefont
  {P.}~\bibnamefont {Wass}}, \ and\ \bibinfo {author} {\bibfnamefont
  {W.}~\bibnamefont {Weber}},\ }\href {\doibase
  https://doi.org/10.1016/j.physleta.2010.06.041} {\bibfield  {journal}
  {\bibinfo  {journal} {Physics Letters A}\ }\textbf {\bibinfo {volume}
  {374}},\ \bibinfo {pages} {3365} (\bibinfo {year} {2010})}\BibitemShut
  {NoStop}%
\bibitem [{\citenamefont {Martinetz}\ \emph {et~al.}(2018)\citenamefont
  {Martinetz}, \citenamefont {Hornberger},\ and\ \citenamefont
  {Stickler}}]{PhysRevE.97.052112}%
  \BibitemOpen
  \bibfield  {author} {\bibinfo {author} {\bibfnamefont {L.}~\bibnamefont
  {Martinetz}}, \bibinfo {author} {\bibfnamefont {K.}~\bibnamefont
  {Hornberger}}, \ and\ \bibinfo {author} {\bibfnamefont {B.~A.}\ \bibnamefont
  {Stickler}},\ }\href {\doibase 10.1103/PhysRevE.97.052112} {\bibfield
  {journal} {\bibinfo  {journal} {Phys. Rev. E}\ }\textbf {\bibinfo {volume}
  {97}},\ \bibinfo {pages} {052112} (\bibinfo {year} {2018})}\BibitemShut
  {NoStop}%
\bibitem [{\citenamefont {Rolke}\ \emph {et~al.}(2005)\citenamefont {Rolke},
  \citenamefont {López},\ and\ \citenamefont {Conrad}}]{ROLKE2005493}%
  \BibitemOpen
  \bibfield  {author} {\bibinfo {author} {\bibfnamefont {W.~A.}\ \bibnamefont
  {Rolke}}, \bibinfo {author} {\bibfnamefont {A.~M.}\ \bibnamefont {López}}, \
  and\ \bibinfo {author} {\bibfnamefont {J.}~\bibnamefont {Conrad}},\ }\href
  {\doibase https://doi.org/10.1016/j.nima.2005.05.068} {\bibfield  {journal}
  {\bibinfo  {journal} {Nuclear Instruments and Methods in Physics Research
  Section A: Accelerators, Spectrometers, Detectors and Associated Equipment}\
  }\textbf {\bibinfo {volume} {551}},\ \bibinfo {pages} {493} (\bibinfo {year}
  {2005})}\BibitemShut {NoStop}%
\end{thebibliography}%

\end{document}